\documentclass[12pt]{article}

\usepackage{amsfonts, amsmath}
\usepackage[dvips]{graphics}

\newcommand{\be}{\begin{equation}} \newcommand{\ee}{\end{equation}}

\def\diag{\ensuremath{\mathop{\rm{diag}}}}

\begin{document}
\title{Conformally Flat Metric, Position-Dependent Mass and Cold Dark Matter} \thispagestyle{empty}
\author { Lev M. Tomilchik, Vladimir V. Kudryashov \hspace{1.5mm} \thanks {
E-mail: l.tomilchik@dragon.bas-net.by; kudryash@dragon.bas-net.by
} }
\date{}
\maketitle
 \vspace{-25pt}
{\footnotesize\noindent Institute of Physics, National Academy of
Sciences of Belarus,\\
 68 F. Skaryna Ave. 220072 Minsk,
Belarus}\\ {\ttfamily{\footnotesize
\\
\rm\normalsize \vspace{0.5cm}
\begin{abstract}
The maximal acceleration (MA) problem associated with the
position-dependent rest mass concept is considered. New arguments
in favor of the mass-dependent maximal acceleration (MDMA) are
put forward. The hypothesis that there exists a maximal force
with the numerical value equal to the inverse Einstein's
gravitation constant is advanced. The Lagrangian and Hamiltonian
classical dynamics of a point-like particle with the
coordinate-dependent mass is given. The effective Lagrangian for
the pure gravitational interaction of a test particle is proposed.
Within the scope of this model the typical spiral galaxy rotation is
described. It is shown that by this model the peculiar
form of the corresponding rotation curve is as a whole reproduced
without recourse to the dark matter concept. Also, it is demonstrated
that the canonical quantization of this model leads directly to the
Dirac oscillator model for a particle with Plank's mass.
 \end{abstract}

\section{Introduction}
The maximal acceleration hypothesis was first conjectured by
Caianiello \cite{r1}. Different aspects, formulation and
inferences concerning possible existence of the limiting value for
the proper acceleration of a particle have been considered in
various works on the classical and quantum bases (see, for example
\cite{r2} and the references therein). Despite the fact that there
are many arguments supporting the existence of MA, its actual
status is still open to dispute. Specifically, it is not clear
whether the numerical value of MA should be considered as a
universal constant, similar to the speed of light, or as a
parameter depending on the individual mass due to the action of
force (Mass-Dependent Maximal Acceleration - MDMA). It was shown
that the effective conformally flat metric can arise directly from
the existence of maximal acceleration \cite{r3}. This metric
reveals interesting confinement aspects. Namely, the
Lorentz-scalar potential and damping normalization factor
introduced in particular relativistic phenomenological quarkonium
models to provide quark confinement occur  (in this case) purely
geometrically as a consequence of the existence of the effective
conformally flat static metric \cite{r4,r8}. As it was first shown
by A.K.Gorbatsevich and L.M.Tomilchik \cite{r4} (see also
\cite{r5}-\cite{r7}), and somewhat later independently by
M.Gasperini \cite{r8}, such a metric leads to the appearance of
the coordinate-dependent rest mass. The position-dependent mass
was introduced by several authors in nonrelativistic quantum
mechanical models developed to study the electronic properties of
condensed media (see, for example \cite{r9}), including rather
interesting attempts of pure geometrical interpretation of such a
dependence via the constant curvature space \cite{r10}.

The apparent efficiency of the quantum-mechanical models based on
the {\it ad hoc} hypothesis of a coordinate-dependent rest mass
suggests that this concept has universal nature. Therefore, it
seems natural to extend it to the classical (nonquantum) objects
as well as to find the general principles that could naturally
lead to such a dependence. It is likely that a search for
kinematic restrictions involving the maximal-acceleration
hypothesis has considerable promise.

In any case, it is natural to expect that the observable effects
possible due to the existence of MA  should appear already at
a level of the point-like particle classical dynamics.

In the present paper new arguments in favor of MDMA are put
forward. The Lagrangian and Hamiltonian classical dynamics of a
point-like particle with the coordinate-dependent mass is
considered. Under the special choice of such dependence, the
effective Lagrangian for the pure gravitational interaction is
proposed. It is shown that within this model the peculiar form of
the corresponding rotation curve is as a whole reproduced without
the use of the cold dark matter concept. It is demonstrated that
the canonical quantization of this model leads directly to the
Dirac oscillator model for a particle with Plank's mass.

\section{Some Heuristic Considerations}

It is well known that the existence of a maximal transmission velocity
for the signal synchronizing clocks separated spatially
in each fixed inertial reference frame leads to the appearance
of four-dimensional space-time with the pseudo-Euclidean
structure.

But according to Einstein and Poincare, the procedure of clock
synchronization suggests the transmission of an instantaneous signal,
i.e. a signal whose duration can be made as small as is wished. On the
other hand, a synchronizing signal can be transmitted and accepted
only due to the exchange of any finite portion of energy $\delta E$ (and
hence momentum $\delta p$). If $E$ and $p$ are certain time-dependent
functions, we have evident relations
\\
$$\delta E=\frac{dE}{dt}\delta t,\delta p=\frac{dp}{dt}\delta t$$
\\
relating the formation duration of a synchronizing signal
($\delta t$) to the values of the energy and momentum carried by
the transmitted signal. In principle, the quantities $\delta E$ and
$\delta p$ should be finite, whereas the interval  $\delta t$
should always tend to zero, implicitly suggesting
fulfillment of the conditions
\begin{equation}\label{U1}
\lim\limits_{\delta t\rightarrow
0}\frac{dE}{dt}=\lim\limits_{\delta t\rightarrow
0}\frac{dp}{dt}=\infty.
\end{equation}
However, if the quantities $(\frac{dE}{dt})_{lim}$ and
$(\frac{dp}{dt})_{lim}$, respectively representing maximal power and
maximal force, are assumed to be limiting, a nonzero extra
time retardation occurs:
\\
$$\delta t_{extra}=(\frac{dE}{dt})^{-1}_{lim}\delta E,\enskip
(or\enskip
 (\frac{dp}{dt})^{-1}_{lim}\delta p)$$
\\
that should be taken into account in clock synchronization. It is
clear that the smaller the distance between the synchronized
clocks the more evident the corrections (recall that the Einstein
clock is point-like by definition). Basing on this reasoning, it
is suggested that there exists a constant $\kappa_{0}$
representing the upper limit for possible values of the proper
energy changing with time \cite{r11}. In this paper, it has been
proposed to combine the infinitesimal intervals of the Minkowski
space $ds$ and momentum space $dp$ in accordance with the Born
Reciprocity Principle \cite{r12}:
\begin{equation}\label{U2}
dS^{2}=ds^{2}+\frac{1}{\kappa_{0}^{2}}dp^{2}=dx^{\mu}dx_{\mu}+
\frac{1}{\kappa_{0}^{2}}dp^{\mu}dp_{\mu}.
\end{equation}
Using the customary definition $ds=c d\tau$ ($\tau$ is the proper
time), we obtain
\begin{equation}\label{U3}
dS^{2}=ds^{2}\{1+\frac{1}{F^{2}_{0}} \frac{dp^{\mu}}{d\tau}
\frac{dp_{\mu}}{d\tau}\}
\end{equation}
where the parameter $F_{0}=\kappa_{0}c$  is a constant with the
dimension of force.  The model proposed by Caianiello and his
coworkers (see \cite{r1} ,\cite{r2}, \cite{r3}) to include the
effects of maximal acceleration into the particle dynamics
consisted in extension of the space - time manifold to the
eight-dimensional space - time tangent bundle.

The fundamental infinitesimal interval for a particle is represented
by the following eight-dimensional line element:
\begin{equation}\label{U4}
dS^{2}= dx^{\mu}dx_{\mu}+\frac{c^{2}}{A^{2}}d \dot x^{\mu}d \dot
x_{\mu}
\end{equation}
\noindent where $\dot x_{\mu} = \frac{dx_{\mu}}{d\tau}$. $A$ is a
parameter with the dimension of acceleration. Assuming the
Minkowski background metric, i.e. $g_{\mu \nu}=\eta_{\mu
\nu}=\diag(1,-1, -1, -1)$, and taking into account that $d\dot
x^{\mu}=d\ddot x^{\mu}d\tau$, we obtain from (\ref{U4}) that
\begin{equation}\label{U5}
dS^{2}= \biggl(1+\frac{\ddot x^{\mu} \ddot
x_{\mu}}{A^{2}}\biggr)ds^{2}.
\end{equation}
Here $\ddot x^{\mu}=\frac{d^{2}x^{\mu}}{d\tau^{2}}$, $ds^{2}
=dx^{\mu}dx_{\mu}=c^{2}d\tau^{2}$, and $\ddot x^{\mu}$  is the
space-like vector (i.e. $\ddot x^{\mu}\ddot x_{\mu}<0$). The
explicit form of $\ddot x^{\mu} \ddot x_{\mu}$ in the noncovariant
notation is
\begin{equation}\label{U6}
\ddot x^{\mu}\ddot x_{\mu}=-\gamma^{3}\biggl\{\underline{W}^{2}-
\frac{1}{c^{2}}(\underline{W},\underline{V})^{2}\biggr\}
\end{equation}
where
$\underline{\beta}=\frac{d\underline{r}}{cdt}=\frac{\underline{\dot
r}}{c}$,
$\underline{W}=\frac{d^{2}\underline{r}}{d^{2}t}=\underline{\ddot
r}$, $\gamma=(1-\underline{\beta}^{2})^{-\frac{1}{2}}$ . In case
of $\dot r=0$ we obtain from (\ref{U5}), (\ref{U6}) the formula
\begin{equation}\label{U7}
dS^{2}=c^{2}d\tau^{2}\biggl(1-\frac{\underline{W}^{2}}{A^{2}}\biggr)
\end{equation}
demonstrating the limiting role of $A$.  Taking the interval
(\ref{U3}) for a material point, i.e. assuming
$\frac{dE}{dt}=\frac{1}{c}
\biggl(\underline{V}\frac{d\underline{p}}{dt}\biggr)$, we obtain
the expression
\begin{equation}\label{U8}
dS^{2}=ds^{2}\biggl\{1-\biggl(1-\frac{\underline{\dot{r}}
^{2}}{c^{2}}\biggr)^{-1}
\frac{1}{F^{2}_{o}}\biggl(\frac{d\underline{p}}{dt}\biggr)^{2}
\biggl(1-\frac{\underline{\dot
{r}}^{2}}{c^{2}}\cos^{2}\phi\biggr)\biggr\}
\end{equation}
where $\phi$ is an angle between $\underline{\dot r}$ and
$\underline{\dot p}=\frac{d\underline{p}}{dt}$. From this formula
it follows that using the same assumptions as in the derivation of
relation  (\ref{U7}) and with
$\frac{d\underline{p}}{dt}=\underline{f}$ we obtain the expression
\begin{equation}\label{U9}
dS=cd\tau\biggl(1-\frac{f^{2}}{F^{2}_{o}}\biggr)^{1/2}
\end{equation}
from whence the constant  $F_{o}$ plays the limiting role. It is
natural to identify this constant as maximal force (MF). We assume
that this constant has purely classical nature, i.e. it is not
related to a minimum value of the action $a_{min}= \hbar$ and
represents the inverse of the Einstein gravitation constant, being
numerically determined as
\begin{equation}\label{U10}
F_{o}=\frac{c^{4}}{G}.
\end{equation}
Postulating this universal constant as $\kappa_{o}
(F_{o}=c\kappa_{o})$, we come to the representation of the
mass-dependent maximal acceleration (MDMA)
\begin{equation}\label{U11}
\biggl(\frac{dV}{dt}\biggr)_{max}=W_{max}=A=\frac{F_{o}}{m}=
\frac{c^{4}}{mG}=\frac{2c^{2}}{r_{g}(m)}
\end{equation}
where  $r_{g}(m)=\frac{2mG}{c^{2}}$  is the gravitation radius
corresponding to the (point-like!) mass  $m$.

It is seen from (\ref{U10}) and (\ref{U11} that MDMA has an
obvious classical, Newtonian meaning of the centripetal
acceleration of a test point-like particle rotating uniformly in a
circle whose radius is equal to the "radius" of the Schwarzschild
sphere.

Needless to say that this descriptive pattern should not be
considered literally. Nevertheless, the idea that the
Schwarzschild parameter $r_{g}$  is related to the MDMA scheme
holds much promise. A model for hyperbolic motion of the
point-like mass in the Special Relativity also indicates the
existence of this relation. Here, as it is known, the
corresponding world line is given by the equation
$\underline{r}^{2}-c^{2}t^{2}=r_{0}^{2}$, where $r_{0}$ is a fixed
parameter having the dimension of length. If the initial velocity
is equal to zero, for the absolute value of the three-dimensional
acceleration ${\underline W}$   we obtain the expression
$W=\dfrac{c^{2}}{r_{0}}\left( 1+\left( \dfrac{ct}{r_{0}}\right)
^{2}\right) ^{-\frac{3}{2}}$,  from which it is seen that the
quantity $W_{0}=\dfrac{c^{2}}{r_{0}}$ represents acceleration at
the initial instant of time, i.e. it is the MDMA for the given
mass $m$. If $W_{0}$ is determined by formula (\ref{U11}), the
parameter $r_{0}$  is described as
$r_{0}=\dfrac{mG}{c^{2}}=\frac{1}{2}r_{g}$.

It may be inferred intuitively that, because of the existence of
maximal acceleration, deviations of the mechanical motion of a
point-like particle from the standard dynamics should manifest
themselves under conditions when its acceleration tends to $W_{0}$
(or is comparable to it). It is clear that in case when $A$ is
determined by expression (\ref{U11}), the "kinematic"
manifestations of MDMA will be more evident for the material
points with greater mass. Let us estimate the quantity
$\frac{W}{A}$ for a case of electromagnetic interactions. To this
end, we use the expression for the force of the static Coulomb
interaction of two charges $e$ and  $Ze$ (i. e.
$mW=\frac{e^{2}Z}{r^{2}}$ ). In this case the expression is as
follows:
\\
$$\frac{W}{A}=\frac{f}{F_{o}}=\frac{e^{2}}{r^{2}}\frac{Z}{F_{o}}
\sim\frac{e^{2}G}{c^{4}}\frac{Z}{r^{2}}=\frac{r_{o}^{2}}{r^{2}}$$
\\where $r^{2}_{o}=\frac{e^{2}GZ}{C^{4}}$
and hence the corrections to unity in  (\ref{U9}) will be
equal to  $\sim \biggl(\frac{r_{o}}{r}\biggl)^{4}$.

As it is seen, $r^{2}_{o} \sim \frac{1}{2}r_{c}r_{g}$, where
$r_{c}=\frac{e^{2}}{mc^{2}}$, $r_{g}=\frac{2mG}{c^{2}}$ are the
classical and gravitation radii of the point-like charged mass for
the interaction of two  equal charges $å$.

For all the elementary particles, the quantity $r_{o}$ vanishes,
being equal to $r_{o} \sim (10^{-68})^{\frac{1}{2}} \sim 10^{-34}$
{\it cm}.

Needless to say that such reasoning gives only a rough estimate,
since the use of the classical parameters of electromagnetic
interaction is inadequate in real situation ( it is well known
that for such interactions the quantum effects become significant
at distances on the order of the atomic dimensions). Nevertheless,
a fixed value of the parameter $r_{0}$ ($r_{0}$ is of the order of
Plank's length) points to the fact that the effects related to
MDMA can influence the known elementary particles only at
distances comparable to Plank's length (i.e. at energies of $\sim
10^{19}$Gev) associated with the Plank energy scales. The same may
be valid for strong interactions too.

A distinct situation is observed in case of gravitation
interactions.

First, note that the concepts of force and acceleration are used
in modern physics only in the classical context. This being so, it
is not imperative, in our opinion, to relate the numerical value
of maximal acceleration to the constant $\hbar$, as this is made
commonly (see \cite{r2}). The existence of the maximal force
interpreted purely classically, however, should exclude a fall at
the center for the pattern of mutual attraction of two point-like
particles, or should lead to the appearance of some effective
repulsion. From this standpoint, the situation is qualitatively
similar to the well-known effects of quantum dynamics caused by
the Plank's quantum of action and hence noncommutativity of the
canonically conjugate coordinates and momenta.

To illustrate the "repulsion" effect caused by the classical
maximal force, we use the elementary concepts based on the
Newtonian law of gravitation. If there exists the above-postulated
maximal force, there should exist a minimum distance $r_{0}$
between two attracting point-like masses. This distance may be
determined from the condition
\begin{equation}\label{UIns1}
\dfrac{mMG}{r_{0}^{2}}=\dfrac{c^{4}}{G}
\end{equation}
whence the expression for parameter $r_{0}$ may be found
\begin{equation}\label{UIns2}
r_{0}^{2}=\frac{1}{4}r_{g}R_{g}
\end{equation}
where $r_{g},R_{g}$ are the corresponding Schwarzschild radii.
\\
This result correlates well with the conclusion that the
Schwarzschild sphere is principally impenetrable for the test
classical particle obtained in \cite{r13} on the basis of the
solution of the motion problem in the Schwarzschild field with
regard to maximal acceleration.

It is obvious that the absence of fall at the center in such a
two-particle problem means that there exists some nonzero angular
momentum. Actually, from the standpoint of a "naive" model, a
minimum distance, where the centers of two small balls with masses
$m$ and $M$ and hence radii
$r_{g}=\frac{2mG}{c^{2}},R_{g}=\frac{2MG}{c^{2}}$ can come of each
other, is equal to $r_{g}+R_{g}$. We easily calculate that the
moment of inertia of such a system with respect to its center of
mass (disregarding the proper rotation of the balls) is equal to
\begin{equation}\label{U13S}
I_{0}=(m+M)r_{g}R_{g} .
\end{equation}
For such a system there should exist a maximal frequency of
rotation around the center of mass. An approximate estimation of
the quantity $\omega_{max}$ within the scope of the model
considered gives the expression
\begin{equation}\label{U14S}
\omega_{max}=\frac{c}{R_{g}}.
\end{equation}
Proceeding from this expression, for a minimal value of the angular
momentum we obtain
\begin{equation}\label{U15S}
L_{min}=I_{0}\omega_{max} .
\end{equation}
Actually, this nonzero angular momentum cannot be attributed to
either of the two particles but belongs to the system as a whole.
The situation is similar to Extra Spin in the electric charge -
magnetic monopole system.

Now consider the applicability of the model under study to
classical systems. Apart from such a fairly perfect classical
theory of gravitation as the Standard General Relativity, there is
quite a number of purely gravitational, comparatively isolated
systems, the observed mechanical properties of which may be
described in reality on the basis of the models correlating with
the ordinary Newton approximation. First, consider rotation of the
most abundant, typical spiral galaxies, in a sense,  belonging  to
the simplest astrophysical   objects. There is a reason to believe
that such systems may be considered within the framework  of the
Newton approximation: the observed intragalaxy velocities do not
exceed several thousandth of the speed of light, and the
corresponding Schwarzschild radius measures fraction of one parsec
(this value is  by a factor of $10^{4} - 10^{5}$ smaller than the
characteristic dimensions of a galaxy, even having a mass on the
order of $10^{12}$ of the Sun mass). Therefore, modeling of the
observed rotation of stars around the galactic center by the
nonrelativistic movement of a material point in a
centrally-symmetric field, representing a combination of the
Newton attraction and gravitational attraction potential linearly
increasing with  the distance, seems to be wholly warranted.

At the same time, the observed asymptotic behavior of the
rotational curves for the typical spiral galaxies is in drastic
contradiction with such a theoretical representation. The most
popular idea that should eliminate this contradiction is currently
associated with the cold-dark-matter concept. Alternative
explanation schemes based on the attempts to modify the Newton
model of gravitational interactions (see \cite{r14} and the
literature therein) are also available. In our opinion, using the
coordinate dependence of the rest mass offers great promise here.
As it will be shown below, the use of a simple classical model
enables one to reproduce the general shape of the rotational curve
for the typical spiral galaxy practically over the whole range of
distances from its center.

When using the conception of phase space and Hamiltonian dynamics,
it is convenient to introduce, {\it a la} M.Born \cite{r12}, the
parameters $q_{0}$, $p_{0}$ having the dimensions of length and
momentum, respectively. For each specific physical system with a
finite action these parameters are assumed to be determined by the
relations $q_{0}p_{0}=a, p_{0}/q_{0}=\kappa_{0}$ ,where $\kappa
_{0}$ is a universal constant,($\kappa _{0}=\dfrac{c^{3}}{G}$),and
$a$ is the parameter with the dimension of action determined for
each classical or quantum system. Its minimum value is equal to
the Planck universal constant $\hbar$. Thus, the following
definitions are true for the parameters: $q_{0}=(a\kappa
_{0}^{-1})^{\frac{1}{2}},p_{0}=(a\kappa _{0})^{\frac{1}{2}}$.

It is evident that for $a_{min}=\hbar$ the parameters $q_{0}$,
$p_{0}$ are equal to the corresponding Planck's quantities
$l_{p}$, $p_{p}$. And the ratio $\dfrac{p_{0}}{q_{0}}$ is
independent of $a$, having the same value both for classical and
quantum systems \footnote {Note that from the geometrical
standpoint, the constant $\kappa_{0}=p_{0}/q_{0}$ determines
maximal deformation ("Prokrust strain") of a given phase area
(including an elementary phase cell).}.

Let us consider the relation between maximal acceleration and
conformal transformation.

It is well known that attempts to interpret the special conformal
transformation (SCT) in the Minkowski Space were made even by L.Page
and N.I.Adams  (see \cite{r15} and papers cited there). This
transformation may be written in the following form:
\begin{equation}\label{U4.1}
x'^{\mu}=\sigma(x,b)(x^{\mu}+b^{\mu}x^{2})
\end{equation}
where
 $$
 \sigma(x,b)=(1+2bx+b^{2}x^{2})^{-1},
$$ $$
 bx=b^{\mu}x_{\mu}=b_{\mu}x^{\mu},
$$ $b^{\mu}$ is the four-vector parameter with the dimension of
$(length)^{-1}$. The parameter   $b^{\mu}$ is traditionally
related to the constant relative four-dimensional acceleration of
the reference frame.

Using this interpretation, we write $b^{\mu}=c^{-2}A^{\mu}$ ,
where   $A^{\mu}$ denotes the constant relative  four-dimensional
acceleration. Besides, it is required that the obtained expression
be coincident, in the limit, with the path formula for the
uniformly accelerated movement along the  $x$ axis. The vector
$b^{\mu}$ should be space-like.

Let us write $b^{\mu}$ in the following form:
\begin{equation}\label{U4.3}
b^{\mu}=\{0,c^{-2}W,0,0\}
\end{equation}
where $W$ is the $x$ - component of the three-dimensional acceleration.

For world lines of the point-like particles only the interior of
the light cone is available. Specifically, it is assumed that
$x^{\mu}=\{ct,0,0,0\}$. Then we obtain
\\
 $$ x^{2}=c^{2}t^{2},\; bx=0,\;
b^{2}=-\frac{w^{2}}{4c^{4}},
\sigma(x,b)=1-\biggl(\frac{Wt}{2c}\biggr)^{2}. $$
\\
Consequently, transformations (\ref{U4.1}) take the following form:
\begin{equation}\label{U4.4}
x'^{\mu}=\frac{Wt^{2}}{2}\biggl(1-\biggl(\frac{Wt}{2c}\biggr)^{2}
\biggr)^{-1},\; t'=t\biggl(1-\biggl(\frac{Wt}{2c}\biggr)^{2}
\biggr)^{-1}.
\end{equation}
If the term $\biggl(\frac{Wt}{2c}\biggr)^{2}$ in the denominator
is ignored, as it might be expected, we obtain nonrelativistic
expressions   $x'=\frac{Wt^{2}}{2}$, $t'=t$.

It is seen that higher values of $\frac{Wt}{2}$ are limited by
$c$.

Evidently, the condition $\frac{Wt}{2c}<1$ suggests two variants
of maximal values for the acceleration and corresponding
time interval
 $$
  (a) \quad W_{min}\Delta t_{max}=2c,
 $$
 $$
 (b)
\quad W_{max}\Delta t_{min}=2c .
 $$
It may be conclusively advocated that the existence
of maximal acceleration requires the existence
of a minimal time interval, and conversely if there
is a maximal time interval, there should exist a certain
minimal acceleration.

It seems probable that, in reality, both these opportunities
should be taken into account. In principle, simultaneous existence
of any large and small time intervals is a necessary condition for
any model realization of a physical clock {\it per se}. To
illustrate, consider the circle of a clock dial and the primes on
it as well as, in the general case, large and small periods and
the possibility of comparing them to a set of integers.

For a uniform and isotropic model of the Universe as a whole there
are obvious candidates for $\Delta t_{min}$ and $\Delta t_{max}$:
Planck's time $\tau_{p}$ and the Universe age (inverse of the
Hubble constant).

It is interesting that the existence of $\tau_{max}$ (in
combination with the assumption that there exists a maximal force
$F_{o}=\frac{c^{4}}{G}$ ) should lead to the upper limit of the
total action in the metagalaxy (assuming $\tau_{max}=H^{-1}_{o}$)
 $$
S_{max}=F_{o}c\tau^{2}_{max}=\frac{c^{5}}{GH^{2}_{o}}
 $$
that at a given experimental value of  $H_{o}^{-1}\sim 10^{18}s$
results in $a_{max}=10^{95}erg \cdot s$.

Since the quantum of action $a_{min}=\hbar\sim 10^{-27}erg \cdot
s$, the total number of quanta is equal to $$
  N\sim 10^{122}\sim e^{280} .
$$

\section{Classical Dynamics of a Particle with the Position-Dependent Mass}

As it has been shown in \cite{r4} (see also \cite{r5}-\cite{r7}),
a geodesic equation in the conformally flat metric
\begin{equation}\label{U4.5}
g_{\mu \nu}=U^{2}(x)\eta_{\mu \nu}, \eta_{\mu \nu}=\diag(1,-1, -1,
-1)
\end{equation}
for the static case with
$U(x)=U(\underline{r}),\frac{\partial(U^{2})}{\partial t}=0$
may be written in the following form:
\begin{equation}\label{U4.6}
\frac{d\underline{p}}{dt}+\frac{p_{0}^{2}}{2m}grad(U^{2})=0, \quad
\frac{dm}{dt} =0.
\end{equation}
Here $\underline{p}=m\underline{\dot{r}},
m=c^{-1}p_{0}U(\underline{r}) (1 -
(\frac{1}{c}\underline{\dot{r}})^2 )^{-\frac{1}{2}}$ and $p_{0}$
is a parameter with the dimension of momentum. This equation is
formally coincident with a nonrelativistic equation of motion for
a "particle" of "mass" $m$ in a "potential field"
\begin{equation}\label{U4.7}
\underline{f}=-\dfrac{p_{0}^{2}}{2m}grad(U^{2}) .
\end{equation}
The solution of equation (\ref{U4.6}) gives an essentially new
result: the existence of the peculiar parametric dependence of
"mass" $m$ on the initial conditions due to its appearance in
(\ref{U4.6}) as an integral of motion rather than as a numerical
constant. The momentum $\underline{p}$  is related to the energy
$E$ by the standard relation
\begin{equation}\label{U4.8}
E^{2}-c^{2} \underline{p}^{2}=c^{2}p_{0}^{2}U^{2}(\underline{r}),
\end{equation}
demonstrating that the quantity $c^{-1}p_{0}U(\underline{r})$
plays a role of the coordinate-dependent rest mass.

The same result can be obtained using the standard variational
procedure with an action determined by the linear element
\begin{equation}\label{U4.9}
ds=(g_{\mu \nu}dx^{\mu}dx^{\nu})^{1/2}
\end{equation}
that is defined by the metric (\ref{U4.5}). The associated Lagrangian
is of the form
\begin{equation}\label{U18}
L=-cp_{0}U(r)\biggl(1-\frac{\underline{\dot{r}}^{2}}{c^{2}}\biggl)^{1/2}.
\end{equation}
The form of the corresponding Hamiltonian is as follows:
\begin{equation}\label{U19}
H=c(\underline{p}^{2}+p_{0}^{2}U(r)^{2})^{\frac{1}{2}} .
\end{equation}

In centrally symmetric case (apart from the energy $E$)
there exists a vector integral of motion (angular momentum)
\begin{equation}\label{U20}
\underline{M}=\underline{r}\times\underline{p}=Ec^{-2}\underline{L},
\quad \underline{L}=\underline{r}\times\underline{\dot{r}}.
\end{equation}
In this case, equation of motion (\ref{U4.6})
takes the form
\begin{equation}\label{U21}
\underline{\ddot{r}}+\frac{c^{2}}{2}\epsilon^{-2}r^{-1}\frac{dU^{2}}{dr}\underline{r}=0
\end{equation}
where $\epsilon=(cp_{0})^{-1}E = U(r) (1 -
(\frac{1}{c}\underline{\dot{r}})^2 )^{-\frac{1}{2}}$ is the
reduced energy. A motion will be finite when $U(r)$ is an
increasing function of $r$. It is assumed that this function has
no singularities throughout the domain of its definition. Since
equation (\ref{U21}) is formally coincident with the
nonrelativistic dynamics equation for a point-like particle in a
centrally symmetric potential field, the motion is investigated in
a standard way. In this case the radial velocity $\dot{r}$ is
determined by the following expression:
\begin{equation}\label{U22}
\dot{r}^{2} =
c^{2}(1-\frac{U^{2}(r)}{\epsilon^{2}})-\frac{\underline{L}^2}
{r^2}
\end{equation}
where $\epsilon$ and $\underline{L}$  are the integrals of motion
defined above. The turning points are obtained from the equation
$\dot{r}=0$. The motion will be finite if the function $U(r)$ is
such that this equation has two real positive roots and, in
exceptional cases, the trajectories are closed.

Consider a particular model. To this end, we choose the
expression for the function  $U(r)$ of the form
\begin{equation}\label{U22}
U(r)=\left(1+\frac{r^{2}}{R_{0}^{2}}\right)^{1/2}
\end{equation}
where $R_{0}$ is a parameter having the dimension of length. The
reasoning in favor of this choice is heuristic in character:

. the problem is  exactly solvable;

. at small  $r(\ll R_{0})$ the form of the potential corresponding
to the field of a gravitating mass, continuously distributed with a
constant density, is reproduced by $U^2(r)$;

. the required asymptotic behavior of the velocity is provided at large
$r(\gg R_{0})$;

. the model is reciprocally symmetric in the sense of M. Born.

Furthermore, we assume that the Lagrangian contains the constant
$\overline{C}$ representing the asymptotic limit of the velocity
of mechanical motion, for a test particle in the closed system
considered, rather than the speed of light. In other words, we
deal with a model defined by the effective Lagrangian  of the form
\begin{equation}\label{U23}
L_{eff}=-m_{0}\overline{C}^{2}
 \left(1+ \frac{r^2}{R_{0}^2}\right)^{1/2}
\left(1-\frac{\underline{\dot{r}}^{2}}{\overline{C}^{2}}\right)^{1/2}
\end{equation}
where $\overline{C}$ and $R_{0}$ are experimentally determined
parameters. $m_{0}$ is the mass of the test
particle that, as will be shown later, disappears in the final
result. In this case, the condition $\dot{r}=0$ leads to a
biquadratic equation
\begin{equation}\label{U24}
r^{4}-R_{0}^{2}(\epsilon^{2}-1)r^{2} +R_{0}^{2}(\overline{C})^{-2}
\epsilon^{2}\underline{L}^{2}=0
\end{equation}
whose solution determines the semiaxes of the elliptic
trajectories as follows:
\begin{equation}\label{U25}
r_{\pm}=\frac{R_{0}}{\surd2}(\epsilon^{2}-1)^{1/2}
\{1\pm(1-4\underline{L}
^{2}\epsilon^{2}R_{0}^{-2}(\overline{C})^{-2}(\epsilon^{2}-1))^{-2})^{1/2}\}^{1/2}.
\end{equation}
The circular orbits correspond to the equality $r_{+}=r_{-}$,
namely the condition $R_{0}^{2}{\overline{C}}^{2}(\epsilon^{2}-1)^2
=4\epsilon^{2}\underline{L}^{2}$ that, considering $\underline{L}^{2}=\underline{r}^{2}\times\dot{\underline{r}}^{2}$,
leads to an explicit expression relating the velocity
$v=(\underline{\dot{r}}^2)^{1/2}$ of the circular orbital motion
of a point-like particle to the orbital radius
\begin{equation}\label{U26}
v(r)= \overline{C}\left(2+\frac{R_{0}^{2}}{r^{2}}\right)^{-1/2} .
\end{equation}
As can be seen, the function (\ref{U26}) reproduces remarkably
well the general shape of the rotation curve for a spiral galaxy.
Figure 1 shows experimental data characteristic for the rotation
curve of the NGC 3198 galaxy (this figure has been taken from
\cite{r14}).

\begin{figure}[h!]
     \leavevmode
\centering
\includegraphics {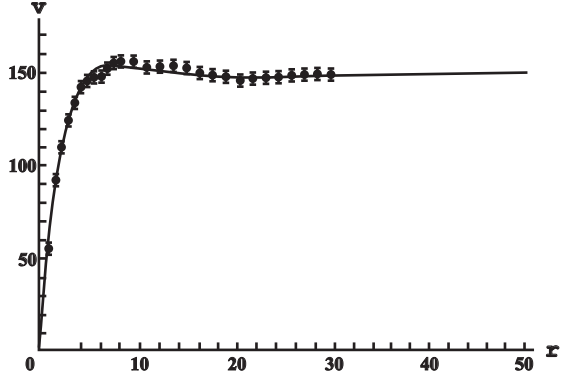}
\caption{Experimental curve of $v(km/s$) as a function of $r(kpc)$ for NGC
3198 galaxy}
\end{figure}

Figure 2 demonstrates the rotation curve $v(r)$ calculated
by formula (\ref{U26}) using the following parameters:
\begin{equation}\label{U27}
\overline{C}= 212.132 km/s, \quad R_{0} = 3.182 kpc .
\end{equation}

\begin{figure}[h!]
     \leavevmode
\centering
\includegraphics {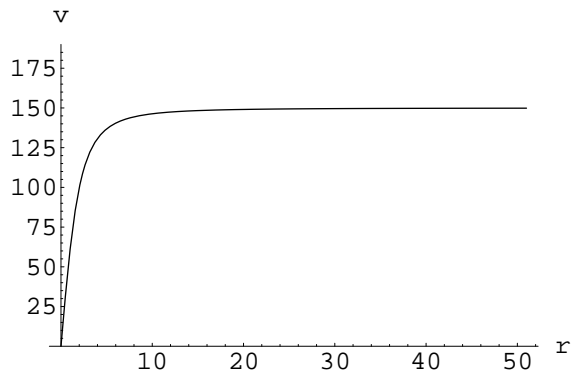}
\caption{Theoretical curve of $v(km/s$) as a function of $r(kpc)$
for NGC 3198 galaxy}
\end{figure}

Note that such a numerical value of the parameter $R_{0}$ seems to
be reasonable as regards the MDMA concept proposed by us. The
problem considered is, to a certain extent, a classical analog of
the quantum oscillator problem. As it is known, here the
characteristic parameter with the dimension of length is
determined as $x_{0}^{2}=\frac{\hbar}{m\omega_{0}}$, where $m$  is
the mass of an oscillating particle and $\omega_{0}$ is the
eigenfrequency of the oscillator. In the case under study the
angular momentum $L_{min}$ determined by formula  (\ref{U15S})
serves as the minimal action. On the other hand, for the quantity
$\omega_{0}$ the following choice seems to be natural in this
case: $\omega_{0}=(\rho_{m}G)^{1/2}$. Here $\rho_{m}$ is a
constant (position-independent) mass density determined as
$\rho_{m}\cong M/R^{3}$, where $M$ is the total mass of the
substance found within a region with the linear dimensions $R$.
Then, for $R_{0}^{2}=\frac{L_{min}}{m\omega_{0}}$, we obtain
accurately to the constant on the order of unity
\begin{equation}\label{U30}
R_{0}=(1+r_{g}/R_{g})^{1/2}(R/R_{g})^{3/4}R_{g}
\end{equation}
where $r_{g}$ and $R_{g}$ are the Schwarzschild radii of a galaxy
and star, respectively, and $R$ is the linear dimension of a
galaxy. For a typical galaxy with $M \sim10^{10}$ mass of the Sun
and $R\sim10^{5}pc$, the value of $R_{0}$ determined from
(\ref{U30}) is about one kpc (kiloparsec).

Note that, in the order of magnitude, the product of the empirical
parameters $\overline{C}^2 R_{0}$ equals to $c^2 R_{g}$, where $c$
is the speed of light and $R_{g}$ is the Schwarzschild radius
corresponding to the galaxy mass. This makes it possible to
suppose that there exists some scale invariance necessitating
special investigation from the standpoint of the
conformally-symmetric dynamics.

\section{Quantization: Dirac Oscillator Model for Plankeon}

Let us show that the quantization procedure based on the function
$U(r)$ of the form (\ref{U22}) leads directly to the well-known
Dirac oscillator model \cite{r16}. In this case the operator
$\widehat{H}^{2}$ corresponding to the Hamilton function
(\ref{U19}) takes the form
\begin{equation}\label{U31}
\widehat{H}^{2}=E_{0}^{2}(\underline{P}^{2}+\underline{\rho}^{2}+1)
\end{equation}
where $P_{k}=-i\frac{\partial}{\partial\xi_{k}},
\rho_{k}=\xi_{k}=\frac{x_{k}}{q_{0}},k=1,2,3$, and
$E_{0}=E_{p}=(c^{5} \hbar G^{-1})^{1/2},q_{0}=l_{p}=(c^{-3} \hbar
G)^{1/2}$ are the corresponding Plank's quantities.

Standard linearization, in accordance with Dirac's procedure,
gives the following expression(in noncovariant notation):
\begin{equation}\label{U32}
\widehat{H}=E_{0}\{(\underline{\widehat{\alpha}}\underline{P})
+\widehat{\beta}(1+\underline{\rho}^{2})^{1/2}\}
\end{equation}
where
$\underline{\widehat{\alpha}},\widehat{\beta}=\widehat{\rho}_{3}$
are the standard  Dirac matrices.

The operator $\widehat{U}^{2}=1+\underline{\rho}^{2}$ suggests
obvious factorization
$\widehat{U}^{2}=\widehat{U}_{+}\widehat{U}_{-}=
\widehat{U}_{-}\widehat{U}_{+}$, where
\begin{equation}\label{U33}
\widehat{U}_{\pm}=1\pm
i(\underline{\widehat{\alpha}}\underline{\rho}) \qquad
(\widehat{U}_{\pm}^{+}=\widehat{U}_{\mp})
\end{equation}
are normal mutually-conjugate Hermitian operators.
In this case
$\widehat{U}_{\pm}=\gamma_{5}\widehat{U}_{\mp}\gamma_{5}$ (in the
given representation,$\gamma_{5}=\widehat{\rho}_{2}$).
Substituting (\ref{U33}) into (\ref{U32}), we obtain two Hermitian
operators
\begin{equation}\label{U34}
\widehat{H}_{\pm}=E_{0}\{(\underline{\widehat{\alpha}}\underline{P})
\pm
i\beta(\underline{\widehat{\alpha}}\underline{\rho})+\widehat{\beta}
\} =E_{0}\{\underline{\widehat{\alpha}}(\underline{P}\mp
i\widehat{\beta}\underline{\rho})+\widehat{\beta}\}.
\end{equation}
As seen, operators (\ref{U34}) are exactly coincident with
Hamiltonian of the Dirac oscillator   $(\widehat{H}_{+})$ and its
supersymmetric partner $(\widehat{H}_{-})$ in the noncovariant
representation (see, for example, \cite{r17}). It is significant
that in the version being considered the model describes a
particle with Plank's mass $m_{p}=(\hbar c G^{-1})^{1/2}$. This
model will be discussed in the context of the gravity quantization
problem in a separate paper.

\section{Concluding Remarks}
In our opinion, the concept of the coordinate-dependent rest mass
(CDRM), along with the hypothesis that there exists the mass-dependent
maximal acceleration (MDMA), may be effectively used
in the field of quantum and classical dynamics.
Gravitational interactions represent an area, where
the models of the classical Lagrangian and Hamiltonian
dynamics with CDRM may be applied. The case in point
is the description of an intermediate region requiring no
recourse to the strict general relativity and making the use
of the Newton approximation insufficient.
There is reason to believe that such an intermediate region
is due to the rotation of large quasi-stationary
cosmological objects, primarily typical spiral galaxies.

It is clear that such phenomenological models should be
substantiated from the standpoint of the standard general
relativity, necessitating special investigation. On the other
hand, the development of such models can help to solve a number of
problems of the relativistic cosmology, e.g., the well-known
singularity problem. It is interesting that the use of the same
phenomenological model makes it possible to give a description of
the behavior of the rotation curves representing the typical
spiral galaxies, in the classical version, and  an exactly
solvable Dirac oscillator model for a spinor particle with Plank's
mass, in the quantum version. It is not improbable that this
enables construction of a theory for the behavior of fermions
against the background of extremely strong gravitation fields.

Note that the existence of the universal constant with the
dimension of momentum/length, postulated by us, allows hyperbolic
geometry to be introduced in each phase plane (and in the
eight-dimensional phase space QTPH) calling for further studies in
subsequent papers.

And, finally, we make some heuristic and philosophical remarks.

Our opinion is that the modern situation with the dark matter is
similar to the situation preceding the development of the special
relativity. At that time, all attempts of elimination of numerous
paradoxes generated by the ether concept based on the dynamic and
ontological principles were unsuccessful. Actually, solution of
the problem has been found by changing the geometry of the
four-dimensional space-time manifold, i.e. owing to a change in
the kinematics. It is probable that the dark matter will repeat
the lot of the ether. We ventured to suggest that the dark matter
is a peculiar factor resultant from the use of inadequate geometry
of the eight-dimensional manifold, representing an extended phase
space, and hence the use of inadequate kinematics.

Section 3 of this paper contains the results obtained jointly by
both coauthors. And "all the remaining is on personal
responsibility of the first author, so that all deficiencies in
the text belong to him "\footnote{We took the liberty to borrow
this phrase from the Introduction to the excellent book
"Relativity: The General Theory" by J.L. Synge (L.M.T.)}.

Unfortunately we were not familiar with  publications by G.W.
Gibbons \cite{r18} and C. Schiller \cite{r19} before July 2005.

\section{Acknowledgements}
The authors are grateful to: Professors E.A.Tolkachev,
E.V.Doktorov, A.K.Gorbatseviñh, Yu.A.Kurochkin, S.Ya.Kilin; Dr's
V.A.Mossolov, J.G. Suarez, A.E.Shalyt-Margolin, Yu.P.Vyblyi for
the support and fruitful discussions on the problems under study.
This work is partially supported by the Belarusian Foundation for
Fundamental Research.


\end{document}